\documentclass[fleqn,usenatbib,useAMS]{mnras}
\usepackage{amsmath, amssymb}
\usepackage[T1]{fontenc}
\usepackage{natbib}
\usepackage{hyperref}
\usepackage{epsfig}
\usepackage{graphicx}
\usepackage{tablefootnote}
\usepackage{caption}
\usepackage{wasysym}

\newcommand{\be}{\begin{equation}}
\newcommand{\ee}{\end{equation}}
\newcommand{\bea}{\begin{eqnarray}}
\newcommand{\eea}{\end{eqnarray}}

\newcommand{\ti}{\times}

\title[3.5 keV Absorption Features in AGN Spectra]{Searches for 3.5 keV Absorption Features in Cluster AGN Spectra}
\author[Joseph P. Conlon]{
Joseph P. Conlon$^{1}$
\\
$^{1}$Rudolf Peierls Centre for Theoretical Physics, 1 Keble Road, Oxford, OX1 3NP, UK\\
}
\date{\today}

\pubyear{2017}

\begin{document}
\label{firstpage}
\pagerange{\pageref{firstpage}--\pageref{lastpage}}
\maketitle

\begin{abstract}
We investigate possible evidence for a spectral dip around 3.5 keV in central cluster AGNs, motivated by previous results
for archival \emph{Chandra} observations of the Perseus cluster and the general interest in novel spectral features around 3.5 keV that may
arise from dark matter physics. We use two deep \emph{Chandra} observations of the Perseus and Virgo clusters that
have recently been made public. In both cases, mild improvements in the fit ($\Delta \chi^2 = 4.2$ and $\Delta \chi^2 = 2.5$) are found by including such a dip at 3.5 keV into the spectrum. A comparable result ($\Delta \chi^2 = 6.5$) is found re-analysing archival on-axis \emph{Chandra} ACIS-S observations of the centre of the Perseus cluster.
\end{abstract}

\section{Introduction}
\label{introduction}

The 3.5 keV line (\cite{Bulbul, Boyarsky}) remains one of the most intriguing potential signals of new physics
from astronomical data. This line was first seen in 2014 in the emission
spectrum of galaxy clusters, both in a stacked sample of distant clusters and also from the individual spectrum of Perseus
and the stacked spectrum of the Coma, Ophiuchus and Centaurus clusters. Since then, there has been an intense observational
effort to investigate first the reality and secondly the origin of this feature, which has resulted in both
further detections and non-detections of this emission line (\cite{Urban:2014yda, Franse:2016dln, Tamura:2014mta, Anderson:2014tza, Riemer-Sorensen:2014yda, Carlson:2014lla, Jeltema:2014qfa, Jeltema:2015mee, Boyarsky:2014ska, Ruchayskiy:2015onc, Iakubovskyi:2015dna, Hofmann:2016urz, Bulbul:2016yop, Hitomi, Cappelluti:2017ywp}).

In terms of possible physical origins for the line, a number of
explanations have been put forward.
These include both instrumental effects and the complex gastrophysics of charged plasmas, the latter including
either potassium or argon thermal emission (\cite{Bulbul, 1411.1758}) and non-thermal sulphur charge exchange (\cite{GuKaastra}). In addition to such more conventional explanations, there is also
the exciting possibility, discussed in many papers, that the line originates from dark matter physics. It is fair to say that there is currently no consensus as to the origin of the line, or even its reality. It had been expected that deep observations with the \emph{Hitomi} Soft X-ray Spectrometer (SXS) would decisively resolve these questions, but sadly the data from \emph{Hitomi} is in practice limited to one deep ($\sim 280$ ks) observation of Perseus with the gate valve on, which for the purpose of the 3.5 keV line corresponds to an effective 70ks exposure at design specification.

All the early studies of the 3.5 keV line only involved the line in emission.
Although this is the only possibility for the case of sterile neutrino dark matter, it would be a mistake to allow this particular model of dark matter to dictate the form of observational searches and interpretations.

More recently, evidence has also appeared for the line in absorption, in particular in the spectrum of NGC1275, the central AGN of the Perseus cluster. In \cite{160501043, 160801684}, analysis of highly off-axis \emph{Chandra} observations of NGC1275 found evidence at around 4 - 5 $\sigma$ for a dip at $3.54 \pm 0.02$ keV.\footnote{The significance is quoted approximately. The precise value of the significance, measured by the $\Delta \chi^2$ improvement when a negative Gaussian is added to the fit, can always be gamed by modifying either the energy range of the spectral fit or the spatial extraction region used for the AGN.} The best-fit strength for this
dip corresponds to an equivalent width of 15 eV (which compares to $\sim 1$ eV for the line in emission).\footnote{For any readers with a particle physics background, the equivalent width of an absorption or emission feature is defined as the width of the continuum spectrum that contains the same number of photons as are present in the feature.}

As described in \cite{160801684}, such a dip would also imply complete consistency between \emph{Hitomi} observations of the centre of the Perseus cluster and previous \emph{Chandra} and \emph{XMM-Newton} analyes of this region (for which the central AGN was excluded from the extraction region). Specifically, as \emph{Hitomi} had much fuzzier angular resolution than either \emph{Chandra} or \emph{XMM-Newton}, it was unable to separately resolve the cluster emission and the AGN emission. The mild ($\sim 2.2 \sigma$) dip present at 3.5 keV in the \emph{Hitomi} data (\cite{Hitomi}) could then be understood as the sum of an excess at 3.5 keV from the diffuse emission and a deficit at 3.5 keV from the AGN spectrum.

If true, the presence of such a strong 3.5 keV line in absorption would radically change the observational situation, as it would be
inconsistent with the vast majority of proposed explanations for the line. In particular, it would appear to require a 2-state dark matter model -- the presence of both absorption and emission requires some analogue of dark atoms to account for both spectral features (a sterile neutrino or other decaying dark matter can only give an emission line), and - besides the absence of a known atomic line at this energy - the equivalent width of the feature is too great to arise from plasma atomic absorption, as the thermal plasma velocity broadening is only $\sim 3 {\rm eV}$. Potential laboratory searches for such a dark matter model are described in \cite{DayFairbairn}.

If real, such a dip must also be universal and not merely a feature of one set of archival observations of the centre of the Perseus cluster. In particular, it must be present in the spectrum of any bright point source shining as a backlight along a large dark matter column density.

What are the ideal observational requirements to observe such a dip? As following the demise of \emph{Hitomi} there are no current imaging microcalorimeter telescopes, we shall assume an energy resolution appropriate to CCD technology (i.e. around 100 eV).\footnote{We do not discuss
gratings observations here, which offer better energy resolution at the price of a loss of imaging capability.} Such a
resolution is then always much larger than the effective width of the dip. What is then necessary to infer the existence of a dip?

There are three critical factors:
\begin{enumerate}
\item
A large number of photons - both the typical power-law spectrum of an AGN and the structure of telescope effective areas imply that the fraction of photons falling within the 3-4 keV band will be relatively small, with most having much smaller energies. As CCD-based telescopes such as \emph{Chandra} or \emph{XMM-Newton} are liable to pileup (resulting in wrong photon energies) when on-axis if more than a single photon arrives from a source within one frame time,
this also requires deep exposures to obtain the large number of required photons.

\item
A large dark matter column density along the line of sight to the source, as the
 strength of any dark matter absorption feature will increase with the amount of absorbing material.

\item
A relatively large velocity dispersion for the dark matter along the line of sight. The ultimate aspect determining
the strength of the signal is the equivalent width of the dip. The size of this is effectively bounded by the dark matter velocity dispersion - as absorption at any energy can never exceed unity, the velocity dispersion effectively bounds the equivalent width, which can only significantly exceed the velocity dispersion if the opacity is exponentially greater than unity.
\end{enumerate}
Taken together, these considerations suggest the optimal targets for any 3.5 keV absorption dip would be bright AGNs that are either in or behind large galaxy clusters. The large photon count comes from the AGN; the cluster screen
provides both the large dark matter column density and the large dark matter velocity dispersion.

Given this, two individual objects stand out for which new deep observations have recently become publicly available. The first is M87, the central AGN of the Virgo cluster, for which 300ks of low frame time reduced pile-up observations became available in June 2017. The second is NGC1275 at the centre of the Perseus cluster, where an additional 100ks of on-axis reduced frame time observations became public in November 2017. We first discuss the Perseus data and then that from M87.

\section{Perseus and NGC1275}

\subsection{Analysis of 2016 data}

Recently 100ks of 2016 observations of the centre of the Perseus cluster became publicly available (obsids 19568, 19913, 19914 and 19915).
Analysis of these datasets is described in \cite{Miller}.
These were taken with ACIS-S and with a reduced frame time of 0.7s. Despite representing only one-tenth the exposure of the 1Ms of on-axis observations of Perseus taken in 2002 - 2004, the more recent data has several advantages.
First, the reduced frame time of 0.7s compared to 3.2s is advantageous in terms of reducing pileup - although as the AGN is approximately four to five times brighter in 2016 than in 2002 - 2004, the actual level of observational pileup is similar. In both cases, this implies that the centre of the image is highly piled and it is necessary to extract from the relatively unpiled wings of the point spread function. Nonetheless, the increased brightness of the AGN implies that the 100ks from 2016 provides as many AGN counts as 500ks from 2002-04, with the increased brightness providing a far greater contrast between the AGN and the background cluster emission than for the older 2002 - 2004 data.

The analysis strategy is as follows. The data was reduced and analysed following the standard \emph{Chandra} pipeline, using
CIAO 4.9 and CALDB 4.7.6.
For each individual observation, we inspected the image using \emph{ds9} and extracted from an annular region around the AGN with inner and outer radius of 0.75" and 3.5" respectively. The outer radius was chosen as the limits of where the AGN emission still dominates that of the cluster itself, while the inner radius is chosen to avoid the highly piled up centre of the image.
The location of the annulus centre was determined from the peak in the counts (we bin in \emph{ds9} to sub-pixel level to determine this as accurately as possible; this is feasible due to the dither of the \emph{Chandra} aimpoint).
For each observation, we extract the spectrum from this annular region, with responses generated assuming a point source
located at the exact centre of the annulus. The background is extracted from an annulus with inner and outer radii 5" and 7".
The four observations were then stacked and the full spectrum analysed, with 41000 counts before background subtraction and 37000 afterwards.

Based on previous data, we are \emph{a priori} interested in a potential deficit around 3.5 keV. To avoid biasing the fit, our strategy is as follows. After subtracting the background, we first exclude the spectral region from 3.3 to 3.7 keV, and fit an absorbed power law from 2.5 to 5.5 keV with this region excluded, using the \emph{chi2datavar} Sherpa statistic. By starting the fit at 2.5 keV, we avoid low-energy contamination from any soft thermal excess associated with the AGN and ensure that emission is dominated by the power law. We fix the absorption at $n_H = 0.2 \ti 10^{22} {\rm cm}^{-2}$, based on the value obtained for our previous fit to the AGN spectrum in \cite{160501043} (we do not directly fit for $n_H$ - a fit starting at 2.5 keV is insensitive to $n_H$ as absorption is relatively unimportant across the energy range of the fit).
We then freeze the parameters of the power law, and re-include the region from 3.3 to 3.7 keV. We now also include a negative
Gaussian at the Perseus redshift ($z=0.0176$) with an intrinsic broadening\footnote{As a reminder, this is not the same as the equivalent width.} frozen at 15 eV (although any value much smaller than the telescope resolution gives an equivalent result). With the power-law parameters now frozen, we allow the energy and amplitude of the negative Gaussian to float (we restrict its energy to between 3 and 4 keV). We report the improvement in the $\chi^2$ that can be achieved by adding
the negative Gaussian to the fit.

\begin{figure*}
  \centering
  \includegraphics[width=0.9\hsize]{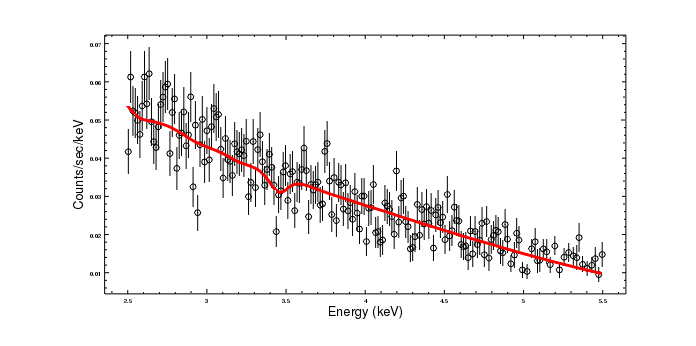}
  \caption{\small The nuclear spectrum of NGC1275 extracted from the 2016 \emph{Chandra} ACIS-S observations, using an annular region with radii 0.75" and 3.5". The power-law fit is shown with an additional negative Gaussian at a best-fit energy of $(3.52 \pm 0.03) {\rm eV}$. }
  \label{NewACIS}
\end{figure*}

Using this method, we find that the negative Gaussian produces an improvement in the $\chi^2$ of 4.2. The best fit energy of the Gaussian is
$(3.52 \pm 0.03) \, {\rm keV}$  and its amplitude $(1.1 \pm 0.5) \ti 10^{-5} \, {\rm ph} \, {\rm cm}^{-2} \, {\rm s}^{-1}$, corresponding to an equivalent width $EW = (16 \pm 7) \, {\rm eV}$. The fitted parameters of the power law are a spectral index of $\Gamma = (2.02 \pm 0.06)$ and a strength of $(8.1 \pm 0.6) \times 10^{-3} \, {\rm ph} \, {\rm cm}^{-2} \, {\rm s}^{-1} \, {\rm keV}^{-1}$ at 1 keV. This value is consistent with the AGN strength measured by $\emph{Hitomi}$ of $9 \times 10^{-3} \, {\rm ph} \, {\rm cm}^{-2} \, {\rm s}^{-1} \, {\rm keV}^{-1}$. The fit is plotted in figure \ref{NewACIS}.

It is important to note here that the accuracy of the amplitude and spectral index for the power law components (which is not the purpose of this paper) is only as good as the accuracy of response matrices for an extraction where the central 0.75" (by radius) of the point spread function is excluded. The systematic errors on the properties of the power law are therefore significantly larger than the statistical fitting errors. It is easy to verify that the fitted power law index becomes progressively harder, reflecting increase pile-up contamination, as the inner annular radius decreases. In the limit where this radius to taken to zero to give a circular extraction, the index reported in \cite{Miller} is $\Gamma = 1.4$.

In contrast, the improvement in the fit due to the negative Gaussian relies only on whether a good fit has been obtained to the
spectral shape of the data, and not on whether these power-law parameters are indeed the true physical parameters of the AGN.
As a local line-like feature, the analysis of such a dip is more robust than finding global properties of the spectrum such as the power-law index.

As a check on the fitting, we also performed a similar analysis where we restricted the data only to the region 3-4 keV (in this case, due to the narrow fitting range we did not first exclude the 3.3 - 3.7 keV region before determining the power law parameters). The results are consistent - inclusion of a negative Gaussian now gave an improvement in the $\chi^2$ of 4.7, with best fit location of $(3.51 \pm 0.03) \, {\rm keV}$ and best-fit amplitude of $(1.3 \pm 0.5) \ti 10^{-5} \, {\rm ph} \, {\rm cm}^{-2} \, {\rm s}^{-1}$.

As a second check, we repeated this analysis extracting from an inner annular radius of 0.5", increasing the number of counts by a third at the cost of a higher pileup fraction.
The best fit energy of a dip now shifts to $(3.55 \pm 0.03) \, {\rm keV}$, with an amplitude of $(9 \pm 4) \ti 10^{-6} \, {\rm ph} \, {\rm cm}^{-2} {\rm s}^{-1}$ corresponding to an equivalent width of $(14 \pm 6) \, {\rm eV}$ and with an improvement in the $\chi^2$ of 5.4.

\subsection{Re-analysis of 2002 - 2004 data}

We also revisit the analysis of the NGC1275 spectrum from the older on-axis ACIS-S observations of the Perseus core that were taken in 2002 and 2004. These were taken on a standard frame time of 3.2s, but at a time when the AGN was much less bright than in 2016 - leading to
a poorer contrast against the cluster emission despite a much deeper exposure than for the 2016 data.
As for the 2016 data, the center of the image is highly piled and it is necessary to exclude this region.
This spectrum was considered in \cite{160501043} from the perspective of constraining axion-like particles (ALPs). However, from the perspective of
the 3.5 keV line, this analysis was sub-optimal.  In particular, we used there large spectral bins that were close to 100 eV in size. This is appropriate to maximise sensitivity for an ALP search, where spectral modulations can spread out over a few hundred eV, but is not optimal for
searches for a narrow line. It is also the case that
in our earlier work of \cite{160501043} we excluded a central square box of side length 2" around the AGN, resulting in a relatively weak contrast against the surrounding cluster emission.

We therefore re-analyse this data in a similar fashion to our treatment of the 2016 data. For each observation, we extract from an annulus of inner radius 0.75" and outer radius 2.5". The smaller outer radius compared to the 2016 data is required as the AGN is dimmer and so it become sub-dominant to the cluster emission earlier in the wings of the PSF. The resulting spectrum has around 90,000 counts in total, which reduces to around 70,000 counts after background subtraction.

\begin{figure*}
  \centering
  \includegraphics[width=0.9\hsize]{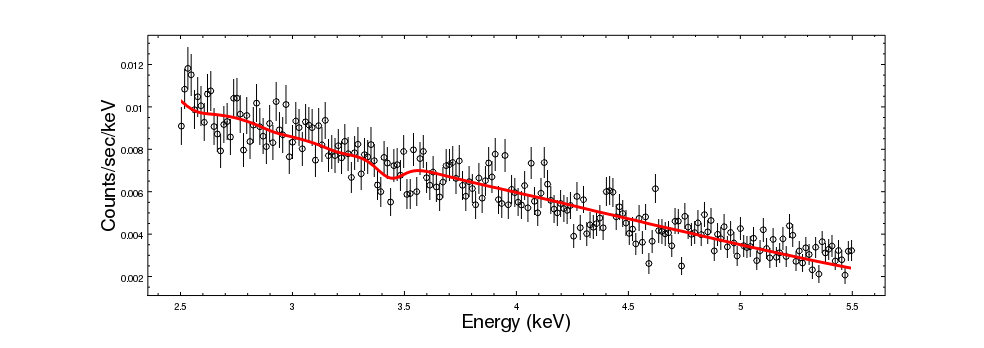}
  \caption{\small The nuclear spectrum of NGC1275 extracted from the 2002 - 2004 \emph{Chandra} ACIS-S observations, using an annular region with radii 0.75" and 2.5". The power-law fit is shown with an additional negative Gaussian at a best-fit energy of $(3.50 \pm 0.04) {\rm eV}$. }
  \label{OldACIS}
\end{figure*}

We follow the same procedure as for the 2016 data. We first group counts to 20 and fit an absorbed power law between 2.5 and 5.5 keV excluding the region between 3.3 and 3.7 keV. The inferred parameters of the AGN power law are a spectral index of $\Gamma = 1.73 \pm 0.04$ and a strength of $(1.40 \pm 0.01) \times 10^{-3} \, {\rm ph} \, {\rm cm}^{-2} \, {\rm s}^{-1} \, {\rm keV}^{-1}$ at 1 keV. Consistent with the brightening of the AGN, this is much weaker than in the 2016 data. We then freeze the power law, re-include the region between 3.3 and 3.7 keV, and add a negative Gaussian to the fit. Allowing the parameters of the Gaussian to float, this improves the fit by $\Delta \chi^2 = 6.5$. The best-fit energy of the Gaussian is $(3.50 \pm 0.04)$ keV with an amplitude of
$(2.4 \pm 1.0) \times 10^{-6} \, {\rm ph} \, {\rm cm}^{-2} \, {\rm s}^{-1}$, which would correspond to an equivalent width of $(14 \pm 6) \, {\rm eV}$. The resulting fit is plotted in figure \ref{OldACIS}.

\section{Virgo and M87}

Recently 300ks of clean data on the M87 nucleus became publicly available
in June 2017 (the use of this dataset to search for ALPs is described in \cite{MarshRussell}). While there are extensive previous observations of M87, what distinguishes these is that the nucleus has recently declined rapidly in brightness, and so the recent data set provides
a deep observation with low levels of pileup.
Reflecting the improved quality of more recent data,
we use all publicly available 2016 and 2017 observations (330ks in total)
but do not include earlier observations as the nucleus was then significantly brighter, with a proportionately higher level of pile-up.

As the M87 nucleus is intrinsically far less bright than that of NGC1275, an annular extraction region is unnecessary and we
extract directly from a circular region around the nucleus. We use a circle of radius 2" centred on the M87 nucleus, and take background from a region that is 5" - 8" from the nucleus, but excluding the jet.

We fit the spectrum with a power law from 2 - 7 keV. As Virgo is at high galactic latitude, the intrinsic absorption is low, with
a galactic $n_H \sim 1 \ti 10^{20} {\rm cm}^{-2}$. As with the Perseus data, this fitting range is rather insensitive
to such values of $n_H$, and so we fix $n_H$
at this value and do not allow it to float.

We follow a very similar approach to the case of the Perseus data. We group counts to 20 and
fit the power law excluding a region from
3.3 to 3.7 keV. The fit has a reduced $\chi^2$ of 0.99 and
results in a power-law index of $\Gamma = 2.30$ and a strength of $5 \times 10^{-4} \, {\rm ph} \, {\rm cm}^{-2} \, {\rm s}^{-1} \, {\rm keV}^{-1}$ at 1 keV. We then freeze the power law parameters
and include into the model a negative Gaussian,
with a redshift set at $z=0.004$ and an intrinsic width set at 15 eV.
We then re-include the signal region and fit for the amplitude and location of the Gaussian, allowing both the amplitude and energy to float.
The inclusion of a Gaussian improves the fit by a $\chi^2$  of 2.5. The best-fit energy of the Gaussian is $(3.50 \pm 0.15)$ keV with an amplitude of $(3.6 \pm 2.2) \times 10^{-7} \, {\rm ph} \, {\rm cm}^{-2} \, {\rm s}^{-1}$, which would correspond to an equivalent width of $(12 \pm 7) {\rm eV}$.

\section{Discussion}

The results we find above are consistent, although at much lower significance,
with our previous results finding evidence for a spectral dip at 3.5 keV in a cluster AGN spectrum (\cite{160501043, 160801684}).
While in these earlier analyses inclusion a dip gave $\Delta \chi^2 \sim 20$, here
the two Perseus spectra each have a preference of $\Delta \chi^2 \sim 4 - 6$ for the inclusion of such a dip,
while the M87 spectrum has $\Delta \chi^2 = 2.5$, which has the right sign but is at a low level of significance.

We also comment about potential contamination to these searches by pileup, given these analyses involve  (of necessity)
bright point sources, which have the potential to be affected
by pileup. This paper is about testing the existence of a localised dip at 3.5 keV in the spectrum of bright AGNs at the centre of galaxy clusters. If this dip is real, then it manifests itself as a local deficit in the number of photons around 3.5 keV. There is no known reason why 3.5 keV should be regarded as a `special' energy in the \emph{Chandra} effective area or instrumental response. Pileup redistributes photons from their true energies to false ones, but there does not appear to be any reason why it should create new features at precisely 3.5 keV.
Severe pileup, where the majority of photons are re-distributed away from their true energies, will wash out any true localised spectral dip.
However, the presence of a small pileup fraction will not affect a true localised dip around 3.5 keV - although the global power-law index or normalisation may be modified, a local absence of photons at 3.5 keV will remain a local absence of photons at 3.5 keV.

\section{Conclusions}

The 3.5 keV line remains one of the most
interesting anomalies in particle astrophysics.
Its origin remains unknown and all methods should be used to determine its nature, unbiased by theoretical preconceptions - which
includes searches for the line in absorption.
The purpose of this paper has been to follow up on our previous results of \cite{160501043} and \cite{160801684} that suggested the presence of an intrinsic dip around 3.5 keV in the spectrum of the central Perseus AGN NGC1275. We have outlined the observational characteristics
that are required to search for such a dip, and have analysed two recently available deep observations of NGC1275 and M87 from this perspective.

In both cases there is a mild improvement in the $\chi^2$ when including an extra negative Gaussian into the fit.
These improvements are not large enough to be especially significant by themselves, but are interesting in the light of the previous results.
Re-analysis of the 1Ms of archival on-axis observations from 2002 - 04 provides a similar picture.

Despite these promising hints,
CCD observations are nonetheless limited in capability and future high-resolution data is required for detailed determination of the nature of the 3.5 keV line.

\section*{Acknowledgements}

I thank Francesca Day, Nick Jennings, Sven Krippendorf and Markus Rummel for discussions.

\bibliographystyle{hapj}
\bibliography{35KevDipBib}

\label{lastpage}
\end{document}